\documentclass[letterpaper,twocolumn,10pt]{article}
\usepackage{usenix2019}
\usepackage{booktabs}
\usepackage{graphicx}

\usepackage[small,bf]{caption}
\usepackage[inline]{enumitem}
\usepackage[compact]{titlesec}

\usepackage{cite,url}
\usepackage{amsmath}
\usepackage{amssymb}

\usepackage{xspace}
\usepackage{subfig}
\usepackage{xcolor}

\usepackage{hyperref}
\hypersetup{
  colorlinks=true,      
  linkcolor=blue,       
  citecolor=magenta,    
  filecolor=cyan,       
  urlcolor=red          
}

\usepackage{tikz}
\newcommand*\circled[1]{\tikz[baseline=(char.base)]{
            \node[shape=circle,draw,inner sep=0pt] (char) {#1};}}

\graphicspath{{figures/}}

\usepackage{algorithm, algorithmicx}
\usepackage[noend]{algpseudocode}
\algrenewcommand\algorithmicindent{0.5em}%
\usepackage{xparse}
\makeatletter
\NewDocumentCommand{\LeftComment}{s m}{%
  \Statex \IfBooleanF{#1}{\hspace*{\ALG@thistlm}}\(\triangleright\) #2}
\makeatother

\newenvironment{denseitemize}{
\begin{itemize}[topsep=2pt, partopsep=0pt, leftmargin=1.5em]
  \setlength{\itemsep}{2pt}
  \setlength{\parskip}{0pt}
  \setlength{\parsep}{0pt}
}{\end{itemize}}

\newenvironment{denseenum}{
\begin{enumerate}[topsep=2pt, partopsep=0pt, leftmargin=1.5em]
  \setlength{\itemsep}{2pt}
  \setlength{\parskip}{0pt}
  \setlength{\parsep}{0pt}
}{\end{enumerate}}

\newtheorem{observation}{Observation}

\def\name{{Salus}\xspace}

\def\ie{{i.e.\xspace}}
\def\eg{{e.g.\xspace}}
\def\etal{{et al.\xspace}}

\def\etc{etc.\xspace}

\begin{document}

\date{}

\title{\bf {\name}: Fine-Grained GPU Sharing Primitives \\for Deep Learning Applications}

\author{
    {\rm Peifeng Yu} \\
    peifeng@umich.edu \\
    University of Michigan
\and
    {\rm Mosharaf Chowdhury} \\
    mosharaf@umich.edu \\
    University of Michigan
}

\date{}

\maketitle

\thispagestyle{empty}

\abstract
GPU computing is becoming increasingly more popular
with the proliferation of deep learning (DL) applications.
However, unlike traditional resources such as CPU or the network, modern GPUs do not natively support fine-grained sharing primitives. 
Consequently, implementing common policies such as time sharing and preemption are expensive.
Worse, when a DL application cannot completely use a GPU's resources, the GPU cannot be efficiently shared between multiple applications, leading to GPU underutilization.

We present {\name} to enable two GPU sharing primitives: \emph{fast job switching} and \emph{memory sharing},
in order to achieve fine-grained GPU sharing among multiple DL applications. 
{\name} implements an efficient, consolidated execution service that exposes the GPU to different DL applications,
and enforces fine-grained sharing by performing iteration scheduling and addressing associated memory management issues.
We show that these primitives can then be used to implement flexible sharing policies such as fairness, prioritization, and packing for various use cases.
Our integration of {\name} with TensorFlow and evaluation on popular DL jobs show that {\name} can improve the average completion time of DL training jobs by $3.19\times$, GPU utilization for hyper-parameter tuning by $2.38\times$,
and GPU utilization of DL inference applications by $42\times$ over not sharing the GPU and $7\times$ over NVIDIA MPS with small overhead.

\endabstract

\section{Introduction}

Deep learning (DL) has received ubiquitous adoption in recent years across many data-driven application domains, ranging from machine translation and image captioning to chat bots and personal assistants \cite{deep-learning}. 
Consequently, both industry and academia are building DL solutions -- \eg, TensorFlow \cite{tensorflow}, CNTK \cite{cntk}, Caffe2 \cite{caffe2}, and others \cite{distbelief, torch, caffe, mxnet, singa, pytorch, theano} -- to enable both \emph{training} of DL models using large datasets as well as serving DL models for \emph{inference}.

GPUs have emerged as a popular choice in this context because they excel at highly parallelizable matrix operations common in DL jobs \cite{nvidia-volta, cnnlab, tpu, tensorflow}. 
Unfortunately, the minimum granularity of GPU allocation today is always the entire GPU -- \emph{an application can have multiple GPUs, but each GPU can only be allocated to exactly one application} \cite{tf-issue-4196, tf-issue-9080, mxnet-issue-4018,nvidia-compute-mode}.
While such exclusiveness in accessing a GPU simplifies the hardware design and makes it efficient in the first place, it leads to two major inefficiencies.

First, the coarse-grained, one-at-a-time GPU allocation model hinders the scheduling ability of GPU cluster managers \cite{mesos, kubernetes, yarn, kubeflow,nvidia-compute-mode}.
For flexible scheduling, a cluster manager often has to suspend and resume jobs (\ie, preempt), or even migrate a job to a different host. 
However, a running DL job must be fully purged from the GPU before another one can start, incurring large performance overhead. 
As such, GPU clusters often employ non-preemptive scheduling, such as FIFO~\cite{kubeflow}, which is susceptible to the head-of-line (HOL) blocking problem.

Second, not all DL jobs can fully utilize a GPU all the time (\S\ref{sec:motivation}). On the one hand, DL training jobs are usually considered resource intensive.
But for memory-intensive ones (\eg, with large batch sizes), our
analysis shows that the average GPU memory utilization is often less than 50\% (\S\ref{sec:room} Figure \ref{fig:exp1-all}) due to varied memory usage over time and between iterations. Similar pattern can also be observed in compute-intensive training jobs.
DL model serving also calls for finer-grained GPU sharing and packing.
Because the request rate varies temporally within the day as well as across models,
the ability to hold many DL models on the same GPU when request rates are low can significantly
cut the cost by decreasing the number of GPUs needed in serving clusters \cite{clipper,tensorrt}.

Additionally, the increasingly popular trend of automatic hyper-parameter tuning of DL
models (\eg, AutoML~\cite{hyperopt,hyperband,hyperdrive}) further emphasizes the need to improve GPU utilization. This can be viewed as ``pre-training''. It
is usually done by generating many training jobs in parallel for hyper-parameter exploration, many of which are killed as soon as they are deemed to be of poor quality. Improved GPU utilization by spatiotemporal packing of many of these jobs together results in shorter makespan, which is desirable because of the all-or-nothing property of hyper-parameter exploration jobs -- \ie, the result is useful only after all exploration jobs finish.

To address these issues, we present {\name} to enable fine-grained sharing of individual GPUs with flexible scheduling policies among co-existing, unmodified DL applications.
While simply sharing a GPU may be achievable, doing so in an efficient manner is not trivial (\S\ref{sec:existing-gpu-sharing}).
{\name} achieves this by exposing two GPU sharing primitives: \emph{fast job switching} and \emph{memory sharing} (\S\ref{sec:overview}).
The former ensures that we can quickly switch the current active DL job on a GPU, enabling efficient time sharing and preemption.
The latter ensures high utilization by packing more small DL jobs on the same device.
The unique memory usage pattern of DL applications is the key to why such primitives can be efficiently implemented in \name:
we identify three different memory usage types and apply different management
policies when handling them (\S\ref{sec:timesharing}).
Combining these two primitives together, the fine-grained spatiotemporal sharing can be used to implement a variety of solutions (\S\ref{sec:scheduling}).

We have integrated {\name} with TensorFlow and evaluated it on a collection DL workload consisting of popular DL models (\S\ref{sec:eval}).
Our results show that {\name} improves the average completion time of DL training jobs by $3.19\times$ by efficiently implementing the shortest-remaining-time-first (SRTF) scheduling policy to avoid HOL blocking.
In addition, \name shows $2.38\times$ improvement on GPU utilization for the hyper-parameter tuning workload,
and $42\times$ over not sharing the GPU and $7\times$ over NVIDIA MPS for DL inference applications with small overhead.

\section{Background and Motivation}
\label{sec:motivation}

This section overviews DL jobs' structural characteristics (\S\ref{sec:dl-training}) and analyzes common DL workloads to understand their resource usage patterns and opportunities for GPU sharing (\S\ref{sec:room}).
Later we discuss existing techniques for GPU sharing among them (\S\ref{sec:existing-gpu-sharing}).

\subsection{Deep Learning}
\label{sec:dl-training}
Deep learning (DL) is a class of algorithms that use a stack of nonlinear processing layers to solve common machine learning tasks, \eg, classification, clustering, or prediction \cite{deep-learning}.
A particular layout of such layers forms a \emph{network architecture}, or simply \emph{network}, that is specially designed for domain-specific problems. 
DL networks must be \emph{trained} before they can be deployed for any practical use.
The knowledge gained from the training process is saved to \emph{model parameters}, which are used in addition to input data for the network to compute the final result.
Collectively, the network architecture and the model parameters are called a DL \emph{model}.
Later, the learned model is used to serve \emph{inference} requests.

\paragraph{Forward and Backward Computation}
During inference, the input is propagated through each layer in order to gain the final result. 
This constitutes a \emph{forward pass}. 
During training, an additional \emph{backward pass} is performed, propagating the gradients back while updating model parameters. 
DL training proceeds by alternating between forward and backward passes in an iterative fashion.
The backward pass is often more expensive than the forward one, because it requires more resources to keep all intermediate results produced between layers to compute gradients (\S\ref{sec:room}). 
Both typically involve a large number of matrix operations, leading to the rising popularity of GPU usage in DL workloads.

\subsection{DL Workloads Characteristics}
\label{sec:room}

To understand the resource usage patterns of DL jobs, we analyzed a workload consisting of 15 DL models (Table~\ref{tab:workloads} in Appendix).
The CNNs are from the official TensorFlow CNN benchmarks \cite{tf-cnn-benchmark};
others are selected popular models in respective fields.

In order to cover a wider range of use cases, while keeping the native input characteristics, we varied the batch size to create 45 distinct workloads, as shown in Table~\ref{tab:workloads}.
Note that the batch size specifies the number of inputs (\eg, images for CNNs) trained in each iteration and also affects the size of model parameters.
Thus it has an impact on the time each iteration takes to complete as well as the memory footprint of the model.
Throughout the paper, we uniquely identify a workload by the network name plus the input batch size. For example, \texttt{alexnet\_25} means a job training \texttt{alexnet}, with a batch size of 25.

In terms of GPU resource usage, one can consider two high-level resources: GPU computation resources (primarily in terms of computation time, often 
referred to as GPU utilization in the literture) and GPU memory. 
We found that both are often correlated with the complexity of the DL model. 
However, GPU memory is especially important because \emph{the entire DL model and its associated data must reside in memory for the GPU to perform any computation}; in contrast, computations can be staggered over time given sufficient GPU memory.

In the following, we highlight several key characteristics of GPU memory usage in DL workloads that highlight GPU memory underutilization issues and/or opportunities for improvements.

\begin{figure}[!t]
    \centering
    \includegraphics[width=\columnwidth]{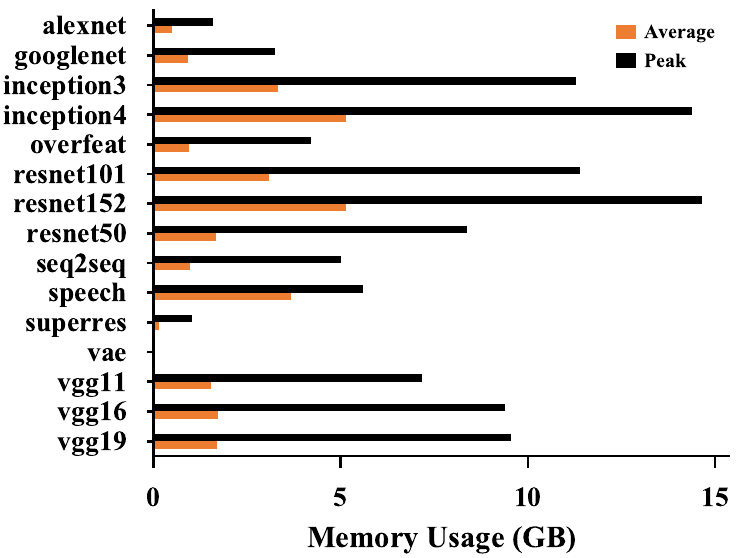}
    \caption{Average and peak GPU memory usage per workload, measured in TensorFlow 
    and running on NVIDIA P100 with 16 GB memory.
    The average and peak usage for vae is 22 MB, 35 MB, which are too small to show in the figure.
    The appendix also includes the measurement in PyTorch (Figure~\ref{fig:mem-pytorch}), which shares a similar pattern.}
    \label{fig:exp1-all}
\end{figure}

\paragraph{Heterogeneous Peak Memory Usage Across Jobs}
DL workloads are known for heavy memory usage \cite{tensorflow, parameter-server, projectadam}. 
Figure~\ref{fig:exp1-all} visualizes the average and peak memory usages of our workloads. 
As networks become larger (with more and wider layers) and the batch size increases, memory requirements of DL jobs increase as well. 
For example, we observed peak memory usages as high as 13.8 GB for \texttt{resnet152} and as low as less than 1 GB for \texttt{vae}.
Such high variations suggest that even during peak allocation periods, it may be possible to run multiple networks on the same GPU instead of running networks in a FIFO manner.

\begin{figure}[!t]
    \centering
    \includegraphics[width=\columnwidth]{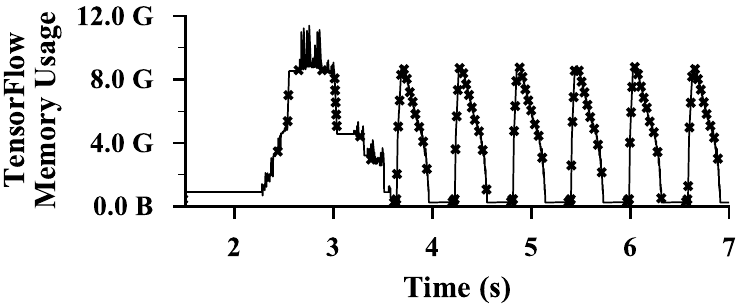}
    \includegraphics[width=\columnwidth]{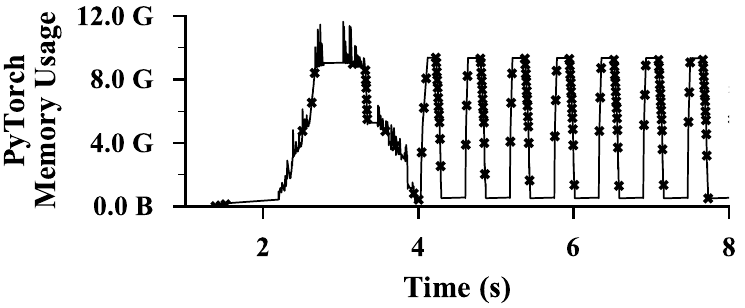}
    \caption{Part of the GPU memory usage trace showing the spatiotemporal pattern when training \texttt{resnet101\_75} and running on NVIDIA P100 with 16 GB memory, using TensorFlow and PyTorch.}
    \label{fig:exp1-tf}
\end{figure}

\paragraph{Temporal Memory Usage Variations Within a Job}
Within each job, however, each iteration of a DL training job is highly predictable with a well-defined peak memory usage and a trough in between iterations. 
Figure~\ref{fig:exp1-tf} shows an example. 
This is because DL jobs go through the same sequence of operations and memory allocations in each iteration. 
The presence of predictable peaks and troughs can help us identify scheduler invocation points.

\paragraph{Low Persistent Memory Usage}
Another important characteristic of GPU memory usage of DL jobs is the use of persistent memory to hold the model of a network -- this corresponds to the consistent troughs across iterations.
Even though the peak usage can be very high, most of it is temporary data created and destroyed within the same iteration.
Fortunately, the size of persistent memory is often very low in comparison to the temporary data, ranging from 110.9 MB for \texttt{googlenet\_25} to 822.2 MB for \texttt{resnet152\_75}.
\emph{As long as the model is already in GPU memory, we can quickly start an iteration of that network.}
This gives us an additional opportunity to improve sharing and utilization.

\subsection{Existing Techniques for Sharing GPUs}
\label{sec:existing-gpu-sharing}
Given that DL workloads leave ample room for GPU sharing, a straw man approach would be disabling the exclusive access mode and statically partitioning (SP) the GPU memory among multiple applications. 
This cannot completely address the underutilization problem due to high peak-to-average memory usage of DL jobs. 
Moreover, static partitioning has significant slowdown compared to the exclusive mode.

NVIDIA's Multi-Process Service (MPS)~\cite{cuda-mps} is the official way to achieve GPU sharing. 
Although users still have to use static partitioning of the GPU memory for each concurrently running job, the performance is better. 
Nonetheless, MPS has limited support for DL frameworks: not all DL framework versions are supported according to our experiments and bug reports on various DL frameworks \cite{tf-issue-4196, tf-issue-9080, mxnet-issue-4018}.
It is possible to achieve GPU memory overcommitting with Unified Memory Access (UMA) \cite{uma}, but it performs poorly due to paging between GPU and the system memory.

A recent work, Gandiva \cite{gandiva}, aims to improve latency and efficiency of DL training by coarse-grained time slicing (\eg, minutes-long slices with about a second switching delay) and static memory partitioning.

NVIDIA's TensorRT Inference server~\cite{tensorrt} achieves simultaneous
DL inference in parallel on one GPU using GPU streams~\cite{gpu-stream}. But it lacks scheduling ability and does not support DL training.

Prior works on fine-grained GPU sharing fall into several categories.
Some attempt to intercept GPU calls -- CUDA calls in the case of NVIDIA GPUs -- and dynamically introduce concurrency by time-slicing kernel execution at runtime \cite{gpgpu-concurrency, cloud-gpu-sharing, chimera}.
Unfortunately, they are either limited to optimizing the efficiency for a single job without considering overall resource utilization or require extensive changes to the underlying GPU architecture.
Others call for new APIs for GPU programming \cite{gloop-mars, pagoda, gnet} but require rewriting existing applications. 
To summarize, these solutions are designed for jobs with a few GPU kernels; as such, they are not scalable to DL applications, where the number of unique kernels can easily goes up to several hundreds.

Table \ref{tab:feature-matrix} summarizes the aforementioned approaches
and a set of desirable features for an ideal solution.

\setlength{\tabcolsep}{2pt}
\begin{table}[!t]
  \small
  \centering
   \begin{tabular}{lcccc}
      \toprule
      Approach           & DL      & Efficiency & Fast          & Flexible          \\
                         & Support &            & Switching     & Scheduling     \\
      \midrule
      Non DL approaches  & No      & -          & -             & -               \\
      SP                 & Yes     & No         & No            & No              \\
      SP + MPS           & Partial & Yes        & Yes           & No              \\
      SP + MPS + UMA     & Partial & No         & Yes           & Yes             \\
      Gandiva            & Yes     & Yes        & No            & No              \\
      TensorRT           & Yes     & Yes        & Yes            & No              \\
      \name              & Yes     & Yes        & Yes           & Yes             \\
      \bottomrule
  \end{tabular}
  \caption{Comparison of GPU sharing approaches (\S\ref{sec:existing-gpu-sharing}).}
  \label{tab:feature-matrix}
\end{table}

\section{{\name}}
\label{sec:overview}

\name \footnote{We have open sourced the system at \url{https://github.com/SymbioticLab/Salus}}
is our attempt to build an ideal solution to GPU sharing.
It is designed to enable efficient, fine-grained GPU sharing while maintaining compatibility with existing frameworks (\S\ref{sec:archi}).
Its overall design is guided by the unique memory usage characteristics of DL jobs.
While existing DL frameworks are limited by the job-exclusive GPU usage scenario, packing multiple jobs onto one GPU changes the combined memory allocation patterns and special care must be taken to mitigate increased fragmentation. 
{\name} addresses both temporal and spatial aspects of the memory management problem by enabling two GPU sharing primitives:
\begin{denseenum}
    \item Fine-grained time sharing via \emph{efficient job switching} among ongoing DL jobs (\S\ref{sec:timesharing});
    \item Dynamic memory sharing via the \emph{GPU lane} abstraction (\S\ref{sec:gpulane}).
\end{denseenum}

Together, these primitives open up new scheduling and resource sharing opportunities. 
Instead of submitting one job at a time, which can easily lead to HOL blocking, one can perform preemption or run multiple DL jobs in a time- or space-shared manner -- all of which can be utilized by a GPU cluster scheduler \cite{gandiva}. 
We demonstrate the possibilities by implementing common scheduling policies such as preempting jobs to implement shortest-remaining-time-first (SRTF), performing fair sharing between jobs, and packing many jobs in a single GPU to increase its utilization (\S\ref{sec:scheduling}).

\subsection{Architectural Overview}
\label{sec:archi}
\begin{figure}[t]
    \centering
    \includegraphics[width=\columnwidth]{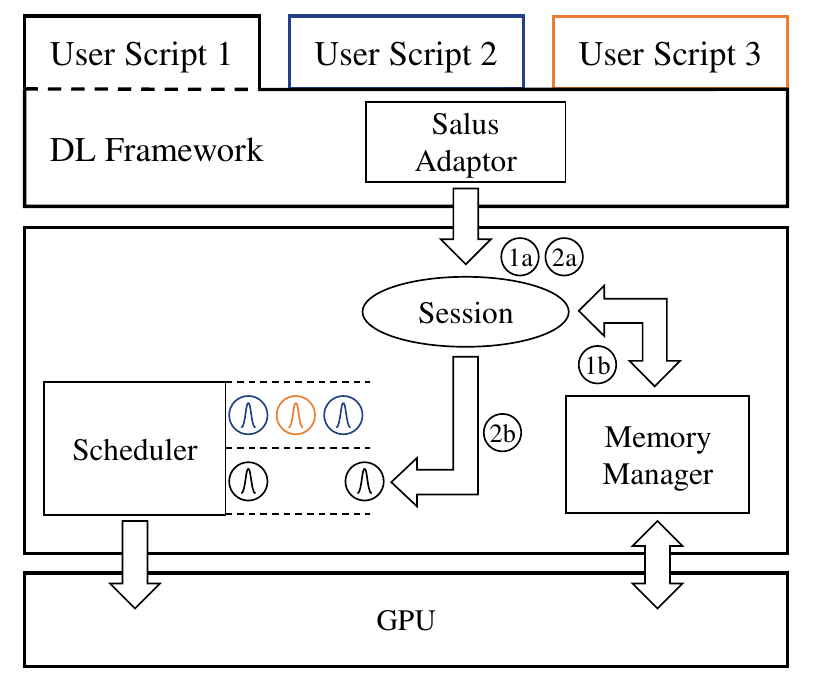}
    \caption{{\name} sits in between DL frameworks and the hardware in the DL 
    stack, being transparent to users.}
    \label{fig:arch}
\end{figure}

At the highest level, {\name} is implemented as a singleton \emph{execution service}, which consolidates all GPU accesses, thus enabling GPU sharing while
avoiding costly context switch among processes on the GPU.
As a result, any unmodified DL job can leverage {\name} using a DL framework-specific \emph{adaptor} (Figure~\ref{fig:arch}).

From a framework's point of view, the adaptor abstracts away low level details,
and \name can be viewed as another (virtual) computation device.

From a user's perspective, the API of the framework does not change at all. 
All their scripts will work the same as they did before.

It is perhaps better to explain the architecture via an example of the life cycle of a
DL job. When a DL job is created in an user script, \emph{\name adaptor} in
the DL framework creates a corresponding session in \name (\circled{1a}).
The computation graph of the DL job is also transferred to \name during the creation.

The session then proceeds to request a lane from the \emph{memory manager} (\circled{1b}). Depending on current jobs in the system, this process can block
and the session will be queued (\S\ref{sec:gpulane}).

During the job's runtime, either training or inferencing, iterations are generated by the user script and forwarded to the corresponding session in \name (\circled{2a}).
They are then scheduled according to their associated GPU lanes by the iteration
scheduler (\circled{2b}), and send to GPU for execution.

The \name execution service thus achieves GPU sharing via iteration-granularity scheduling of DL jobs. We will elaborate on a performance-efficiency tradeoff in choosing this granularity (\S\ref{sec:sched-granularity})

\subsection{Efficient Job Switching}
\label{sec:timesharing}

\begin{figure}
    \includegraphics{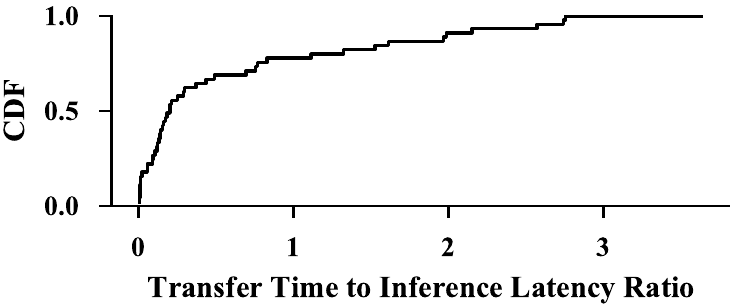}
    \caption{Theoretical minimal transfer time compared with model inference latency. Data collected from our workloads and transfer time is calculated using 30 GB/s for transfer speed.}
    \label{fig:trans-vs-latency}
\end{figure}

The ability to switch between jobs is paramount to implement time sharing and preemption -- two techniques extensively used by modern schedulers in many contexts. 
Suspending a running job and resuming the same or another one have always been possible on GPU as well. 
Modern DL frameworks extensively use checkpointing to mitigate data and computation loss due to the long running nature of DL training jobs.
The same technique is applied by Gandiva \cite{gandiva} to achieve second-scale suspend/resume.
Nevertheless, checkpointing can result in large data transfers from and to the GPU memory, even in the best case when only
model parameters are transfered, the communication time is still non-negligible. It even becomes unacceptable if the system
ever wants to support inference workloads:
the theoretical minimal transfer time can be even several times longer than the
inference latency itself, according to the measurement on our collection of
workloads (Figure~\ref{fig:trans-vs-latency}).

\begin{observation}
Transferring GPU memory back and forth is not practical to achieve low latency
given current GPU communication bandwidth.
\end{observation}

\subsubsection{Characterizing DL Memory Allocations}
We observe that one can push things further by taking a close look at different types of memory allocations in a DL job.
Specifically, we define three types of memory allocations with unique characteristics.
\begin{denseenum}
    \item \emph{Model:} These mostly hold model parameters and typically consist of a few large chunks of memory. 
      They are persistent because they have to be available throughout the whole job's lifetime.
      Because the model size is typically fixed during the entire training process, model data has little or no temporal variations and is predictable. 

    \item \emph{Ephemeral:} These are the scratch memory needed during each iteration. 
      These memory usually hold intermediate layers' outputs as well as temporary data generated by the algorithm itself.
      They are only needed during computations and are released between iterations, giving rise to the temporal memory usage patterns of DL jobs. 
      They are often large memory allocations as well.

    \item \emph{Framework-internal:} These are usually used by the DL framework for book-keeping or for data preparation pipeline.
    They often persist across iterations.
\end{denseenum}

\begin{figure}[t]
    \includegraphics{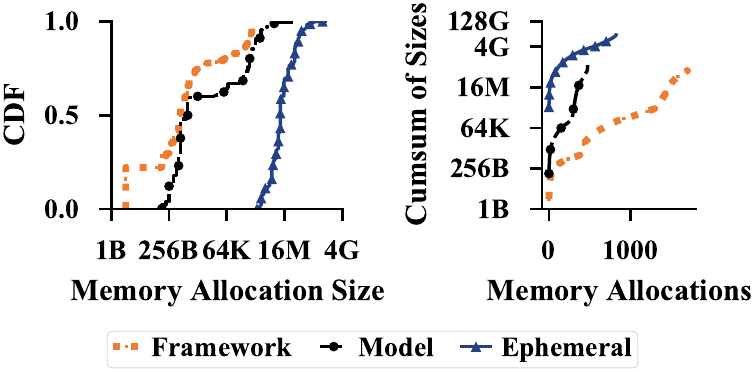}
    \caption{Memory allocation size and number distribution for ephemeral, model and framework-internal memory, using \texttt{inception3\_50} as an example. }
    \label{fig:memory-type-cdf}
\end{figure}

Collectively, model and framework-internal memory are \emph{persistent} across iterations. As an example, Figure~\ref{fig:memory-type-cdf} gives the memory allocation size distribution for a popular CNN workload: \texttt{inception3\_50}.

\begin{observation}
There is significantly less persistent memory usage than ephemeral memory for
a DL job. It is possible to keep more than one job's persistent memory in GPU
while still having enough space for either one's ephemeral memory.
\end{observation}

The above two observations naturally lead to the conclusion that
fast job switching can be enabled by not removing persistent memory from GPU at all.
Thus unlike existing works~\cite{gandiva}, \name is designed to enable significantly faster suspend/resume operations
by keeping persistent memory around,
and then an iteration-granularity job scheduler
(\eg, time-sharing or preemption-based)
decides which job's iteration should be run next.

\subsubsection{Scheduling Granularity}
\label{sec:sched-granularity}
Given that iterations are typically short in DL jobs (ranging from tens of milliseconds to a few seconds), with an even finer granularity, \eg, at the GPU kernel level, it may be possible to further utilize GPU resources. 
However, finer-grained scheduling also adds more overhead to the execution service. 
Indeed, there is a tradeoff between maximum utilization and efficiency for a given scheduling granularity.

To understand this tradeoff, we prototyped a GPU kernel-level switching mechanism as well only to find that scheduling at that level incurs too much overhead for little gain. 
It requires all GPU kernels to go through a central scheduler, which, in addition to becoming a single bottleneck, breaks common efficiency optimizations in DL frameworks such as kernel batching and pipelining. 

\begin{figure}[t]
    \centering
    \includegraphics[scale=1.0]{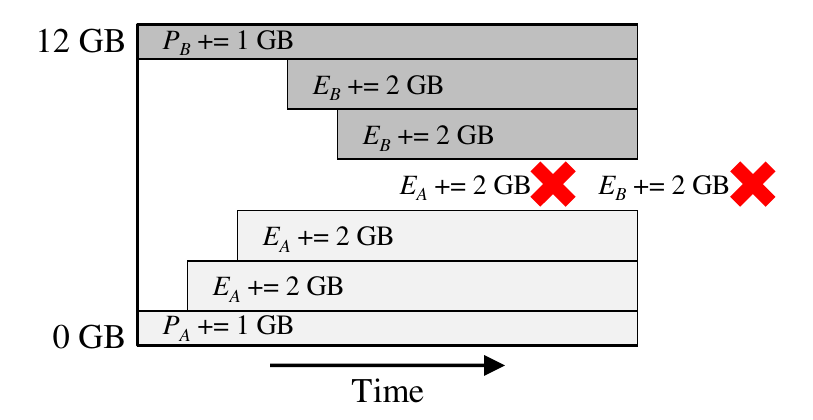}
    \caption{Deadlock from progressive memory allocations.}
    \label{fig:deadlock}
\end{figure}

To make things worse, a unique deadlock issue arises due to the progressive memory allocations performed by many DL frameworks: a job can start its iteration as long as its model memory is available, and then the ephemeral memory is allocated gradually by a series of GPU kernels.
Now consider the following scenario with 12 GB GPU memory capacity, and two iterations from jobs A and B. 
Their model memory usages are $P_A = P_B$ = 1 GB and ephemeral memory usages are $E_A=E_B$ = 7 GB (we are ignoring framework-internal usage because of its relatively smaller size).
Instead of allocating all 8 GB at once, each iteration of a job allocates in different increments.
For example, consider a possible allocation order shown in Figure~\ref{fig:deadlock}, where ($Y_X$ += N GB) refers to job X
allocating N GB of type Y memory.
After a few rounds of successful allocations, if both jobs attempt to allocate their remaining requirements as follows: ($E_A$ += 3 GB) and ($E_B$ += 3 GB), neither will be able to proceed, causing a deadlock!
Mitigating the deadlock would have been simple if GPUs provided program controlable fast paging mechanisms, which unfortunately is not the case today.

In contrast, our choice of switching in between iterations allows us to sidestep the problems of progressive memory allocations.
This is because all ephemeral allocations are released by the framework after each iteration, and model and framework-internal allocations remain constant across iterations.

\subsection{Memory Sharing via GPU Lane}
\label{sec:gpulane}

Although DL jobs' memory usages have spatiotemporal variations, many cannot reach the total capacity of a GPU's memory. 
Naturally, we must consider ways to better utilize the unused memory.

Built on top of the efficient job switching,
we design a special memory layout scheme, the \emph{GPU Lane}, that achieves memory sharing and improves memory utilization.

\begin{figure}
    \includegraphics[width=\linewidth]{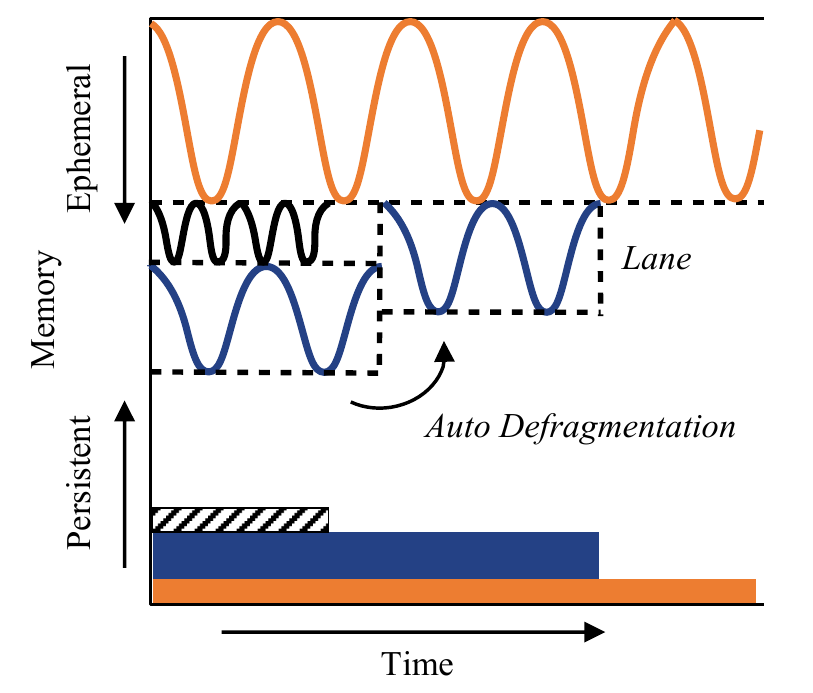}
    \caption{GPU memory is divided into stack and heap, with auto defragmentation at iteration boundaries.}
    \label{fig:lanes}
\end{figure}

First of all, learning from classic memory management techniques of stack and heap to seperate dynamic allocations from static ones, we divide GPU
memory space into \emph{ephemeral} and \emph{persistent} regions,
growing from both end of the memory space (Figure~\ref{fig:lanes}).
A DL job's model and framework-internal memory is allocated in the persistent region,
while its ephemeral memory goes into, obviously, the ephemeral region.

The ephemeral region is further divided into \emph{lanes}, which are continuous
memory spaces that can contain ephemeral memory allocation for iterations.
Lanes are not only about memory, though. Iteration execution is serialized within a lane and parallelism is achieved across lanes, which is implemented using GPU streams. Each lane can be assigned to multiple DL jobs, in which case efficient job switching primitive discussed in previous section is used to time share the lane.

The lane's restriction on execution is necessary because ephemeral memory differs from the other two types of memory in terms of allocation patterns and timing.
As a result, simply putting allocations from two different iterations together
can cause deadlock as already discussed if the memory is oversubscribed.

Even if enough memory is
ensured for both peak memory usage for two iterations, memory fragmentation
can still cause superfluous out-of-memory errors if not handled.
More specifically, while the framework-internal memory allocations are small in
size,
they can have a large impact on the overall memory layout
and may create more memory fragments when multiple iterations are allocating simultaneously.
While there are works implementing a memory planner before actually starting
the iteration~\cite{mxnet}, they are not available to all frameworks.

We approach the problem by first implementing an application-aware bin-based allocator to mitigate the fragmentation. However, it breaks memory optimizations commonly used in DL frameworks because they assume single job running at a time. Since our goal is to fully support existing workloads with minimal impact on the user, we choose to limit the dynamic allocation in the ephemeral region and isolate memory allocations across lanes
to ensure maximum compatibility while achieving adequate flexibility.

\subsubsection{Lane Auto Defragmentation}
Having lanes does not eliminate memory fragmentation, it moves fragmentation within lane to fragmentation at the lane level. However, defragmentation
is much easier at this level. Traditionally, defragmentation is achieved by first moving data out of memory and later moving it back again. In case of lanes, the allocations are released completely at the end of each iteration and goes back at the start of next iteration -- they are ephemeral memory after all. Therefore, defragmentation happens almost automatically at no cost: no extra memory movement is needed.

Consider the situation illustrated in Figure~\ref{fig:lanes}, when the small job stops, its lane space is quickly reclaimed at the iteration boundary by
the job that was allocated below it.

\subsubsection{Lane Assignment}
\label{sec:lane-assign}
It is vital to determine the size and number of lanes in the GPU, as well as
how lanes are assigned to jobs.
\name uses a simple yet efficient algorithm to decide between opening a new lane
and putting jobs into existing lanes.

\begin{algorithm}[tbp]
    \caption{GPU Lane Assignment}
    \label{algo:lane}
    \begin{algorithmic}[1]
        \State $Q \gets \emptyset$ \Comment{The pending queue for new jobs}
        \Procedure{JobArrive}{$P, T$}
            \State $P$: new job's persistent memory requirement
            \State $E$: new job's ephemeral memory requirement
            \State $ Q \gets Q \cup \{(P, E)\}$
            \State \Call{ProcessRequests}{$Q$}
        \EndProcedure
        \Statex
        \Procedure{JobFinish}{$\mathit{lane}$}
            \State $\mathit{lane}$: the lane that the finished job assigned to
            \State $\mathrm{ref}(\mathit{lane}) \gets \mathrm{ref}(\mathit{lane}) - 1$
            \If{$\mathrm{ref}(\mathit{lane}) == 0$}
                \State Delete $lane$
                \State \Call{ProcessRequests}{$Q$}
            \EndIf
        \EndProcedure
        \Statex
        \LeftComment{After a lane is moved due to auto defragmentation}
        \Procedure{LaneMoved}{}
            \State \Call{ProcessRequests}{$Q$}
        \EndProcedure
        \Statex
        \Procedure{ProcessRequests}{$Q$}
            \ForAll{$(P, E) \in Q$}
                \State $\mathit{lane} \gets $ \Call{FindLane}{$P, E$}
                \If{Found $\mathit{lane}$}
                    \State $\mathrm{ref}(\mathit{lane}) \gets \mathrm{ref}(\mathit{lane}) + 1$
                    \State Assign $\mathit{lane}$ to the corresponding job
                \EndIf
            \EndFor
        \EndProcedure
        \Statex
        \Procedure{FindLane}{$P, E$}
            \State $C$: size of total capacity
            \State $P_i$: persistent memory usage of existing job $i$
            \State $L_j$: lane size of existing lane $j$
            \State $\mathbb{L}$: set of existing lanes
            \LeftComment{Try to create a new lane}
            \If{$\sum_i P_i + P + \sum_j L_j + E \le C$}
                \State $\mathit{lane} \gets $ new GPU lane with capacity $E$
                \State \textbf{return} $\mathit{lane}$
            \EndIf
            \LeftComment{Try to put into an existing lane}
            \ForAll{$\mathit{j} \in \mathbb{L}$}
                \If{$L_j \ge E$ and is the best match}
                    \State \textbf{return} $j$
                \EndIf
            \EndFor
            \LeftComment{Try to replace an existing lane}
            \For{$r \in \mathbb{L}$ in $L_r$ ascending order}
                \If{$\sum_i P_i + P + \sum_j L_j - L_r + E \le C$}
                    \State $L_r \gets E$
                    \State \textbf{return} $r$
                \EndIf
            \EndFor
            \State \textbf{return} not found
        \EndProcedure
    \end{algorithmic}
\end{algorithm}

Throughout the process, the following ``safety'' condition is always kept to make sure the persistent region and ephemeral region do not collide into each other:
\begin{align*}
    \sum_{\mathit{jobs}} P_i + \sum_{\mathit{lanes}} L_j &\le C \\
    L_j &= \max_{i \; \mathrm{in} \; j} E_i
\end{align*}
where $P_i$ and $E_i$ are respectively the persistent (model and framework-internal) and ephemeral memory usage of job $i$. $L_j$ is the lane size of lane $j$, which is again defined as the maximum ephemeral memory usage of all jobs in the lane. $C$ is the capacity of the GPU.

By ensuring enough capacity for persistent memory of all the admitted jobs and enough remaining for the iteration with the largest temporary memory requirement, {\name} increases the utilization while making sure that at least one job in the lane can proceed.

At the highest level, the algorithm tries to obtain a lane in the following order, returning once a suitable lane is found and safety condition is met:

\begin{denseitemize}
    \item Open a new lane
    \item Use an existing lane
    \item Reorganize lane assignments to existing jobs to reduce the size of ephemeral region
\end{denseitemize}

As shown in Algorithm~\ref{algo:lane}, the system is event-driven and reacts when there are jobs arriving or finishing, or at iteration boundaries when auto defragmentation happens.

How to reorganize lane assignments is an open question. We find the one implemented in our algorithm works fairly well in practice, but there are more possibilities about finding the optimal number of lanes given a set of jobs.

\section{Scheduling Policies in {\name}}
\label{sec:scheduling}

The state-of-the-art for running multiple DL jobs on a single GPU is simply FIFO -- regardless of the DL framework 
\cite{tf-issue-4196, tf-issue-9080, mxnet-issue-4018} -- that can lead to HOL blocking.
Although Gandiva \cite{gandiva} recently proposed a time sharing approach, it enforces sharing over many minutes because of high switching overhead. 
It uses MPS for memory sharing with admittedly unpredictable performance. 

By enabling fine-grained GPU sharing primitives, \name makes it possible to pack multiple jobs together to increase efficiency, to preempt long-running jobs in favor of shorter ones (or based on other priority criteria), and many others, opening up a huge design space that can 
be explored in future works.

To demonstrate the possibilities, in our current work, we have implemented some simple scheduling policies,
with \name specific constrains (\ie, safety condition).
The \texttt{PACK} policy aims to improve resource utilization and thus makespan, the \texttt{SRTF} policy is an implementation of shortest-remaining-time-first (SRTF), and the \texttt{FAIR} policy tries to equalize resource shares of concurrent jobs.

\subsection{\texttt{PACK} to Maximize Efficiency}
To achieve higher utilization of GPU resources, many jobs with different GPU memory requirements can be packed together in separate GPU lanes based on their peak memory usages. 
However, packing too many lanes exceeding the GPU memory capacity will either crash the jobs or incur costly paging overhead (if UMA is enabled), both of which would do more harm than good.

Consequently, this policy works with ``safety'' condition to ensure that the total peak memory usage across all lanes is smaller than the GPU memory capacity. 
Because each lane has guaranteed resources, there is no fairness consideration in this case. 

Apart from training many different jobs or many hyper-parameter searching jobs in parallel, this can also enable highly efficient inference serving. 
By simultaneously holding many models in the same GPU's memory, we can significantly decrease the GPU requirements of model serving systems like Clipper \cite{clipper}.

\subsection{\texttt{SRTF} to Enable Prioritization}
Developing DL models are often an interactive, trial-and-error process where practitioners go through multiple iterations before finding a good network.
Instead of waiting for an on-going large training to finish, {\name} can enable preemption -- the large job is paused -- to let the smaller one finish faster on the same GPU lane. 
In this way, {\name} can support job priorities based on arbitrary criteria, including based on size and/or duration to implement the shortest-remaining-time-first (SRTF) policy.
The higher priority job is admitted as long as its own safety condition is met -- \ie, at least, it can run alone on the GPU -- regardless of other already-running jobs.

Note that we assume the job execution time is known and thus it is possible to implement SRTF. While there are works on how to estimate such job execution time~\cite{optimus}, the subject is beyond the scope of this paper and we only focus on providing primitives to enable the implementation of such schedulers.

\subsection{\texttt{FAIR} to Equalize Job Progress}
Instead of increasing efficiency or decreasing the average completion time, one may want to time share between many DL jobs during high contention periods \cite{gandiva}. 
Note that there may be many different so-called \emph{fair} algorithms based on time sharing; we demonstrate the feasibility of implementing one or more of them instead of proposing the optimal fairness policy.
Specifically, we admit new jobs into the GPU lane while maintaining the safety condition, and equalize total service over time for jobs in each lane.

\section{Evaluation}
\label{sec:eval}

\begin{figure}[!t]
    \centering
    \includegraphics{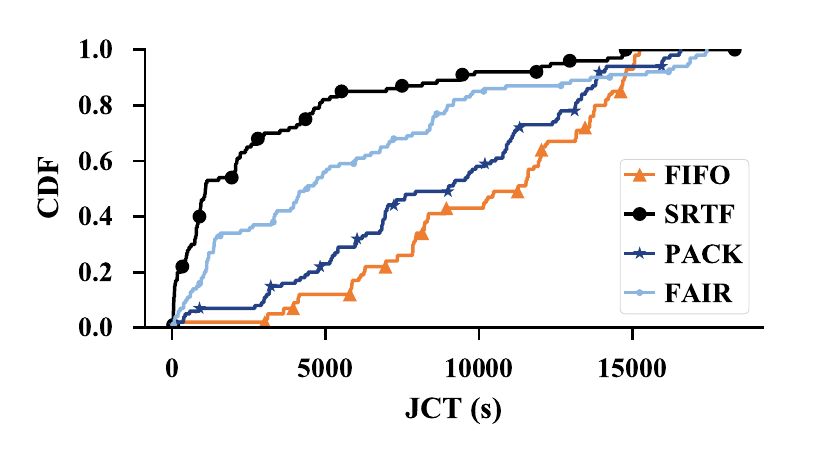}
    \caption{CDFs of JCTs for all four scheduling policies.}
    \label{fig:jct-cdf}
\end{figure}

\begin{table}[!t]
    \centering
    \begin{tabular}{lrrrrr}
        \toprule
        Scheduler &   Makespan & Avg. Queuing & Avg. JCT & 95\% JCT \\
        \midrule
        \texttt{FIFO} & 303.4 min & 167.6 min & 170.6 min & 251.1 min \\
        \texttt{SRTF} & 306.0 min & 28.6 min & 53.4 min & 217.0 min \\
        \texttt{PACK} & 287.4 min & 129.9 min & 145.5 min & 266.1 min \\
        \texttt{FAIR} & 301.6 min & 58.5 min & 96.6 min & 281.2 min \\
        \bottomrule
    \end{tabular}
    \caption{Makespan and aggregate statistics for different schedulers.}
    \label{tab:exp11}
\end{table}

\begin{figure*}[!t]
    \centering
    \subfloat[][A slice of 6 jobs switching between each other. Gray areas represents the waiting between a job arrives and it actually gets to run.]{%
      \label{fig:srtf-compute}
      \includegraphics[scale=1]{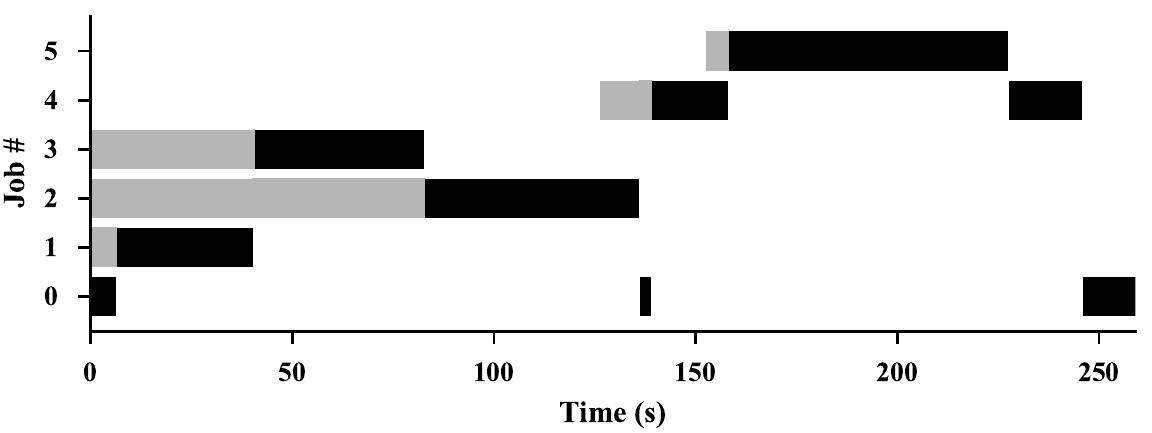}
    }
    \hspace{1cm}
    \subfloat[][Memory usage during a job switching.]{%
      \label{fig:srtf-mem}
      \includegraphics[scale=1]{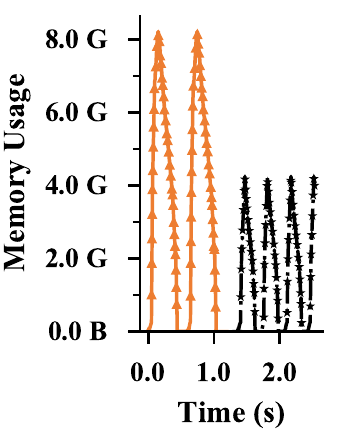}
    }
    \caption{Details of a snapshot during the long trace running with SRTF. 
      In both slices, time is normalized.}
\end{figure*}

We have integrated {\name} with TensorFlow and evaluated it using a collection of training, hyper-parameter tuning, and inference workloads~\cite{tf-cnn-benchmark,seq2seq,vae,superres,deepspeech} to understand its effectiveness and overhead.
The highlights of our evaluation are as follows:
\begin{denseitemize}
    \item \name can be used to implement many popular scheduling algorithms. 
      For example, the preemptive SRTF scheduler implemented in \name can outperform FIFO by $3.19\times$ in terms of the average completion time of DL training jobs (\S\ref{sec:large-scale-experiment}).

    \item Using \name, one can run multiple DL jobs during hyper-parameter tuning stage, increasing GPU utilization by $2.38\times$ (\S\ref{sec:eval-automl}).
    
    \item Similarly, for inference, \name can improve the overall GPU utilization by $42\times$ over not sharing the GPU and $7\times$ over NVIDIA MPS (\S\ref{sec:eval-inference}).

    \item \name has relatively small performance overhead given its flexibility and gains (\S\ref{sec:eval-overheads}).
\end{denseitemize}

\paragraph{Environment}
All experiments were done on a x86\_64 based Intel Xeon E5-2670 machine with 2 NVIDIA Tesla P100 GPUs. 
Each GPU has 16GB on-chip memory. 
TensorFlow v1.5.0 and CUDA 8.0 are used in all cases.

\paragraph{Baseline(s)}
Our primary baseline is the FIFO scheduling commonly used in today's GPU clusters \cite{gandiva}.
We also compare against NVIDIA MPS.

\subsection{Long-Running Training}
\label{sec:large-scale-experiment}
First and foremost, we focus on \name's impact on training.
To this end, we evaluate \name using a job trace of 100 workloads, generated using the jobs described in Table~\ref{tab:workloads}.
We considered multiple batch sizes and durations of each training job in the mix.
The overall distribution followed one found in a production cluster \cite{gandiva}.

We compare four different schedulers:
\begin{denseenum}
  \item \textbf{FIFO} refers to processing jobs in order of their arrival. 
    This is the de facto mechanism in use today in the absense of \name.
    
  \item \textbf{SRTF} is a preemptive shortest-remaining-time-first scheduler.
    We assume that the duration is known or can be estimated using existing techniques~\cite{optimus}.
    
  \item \textbf{PACK} attempts to pack as many jobs as possible in to the GPU. 
    The goal is to minimize the makespan.
  
  \item \textbf{FAIR} uses time sharing to equally share the GPU time among many jobs.
\end{denseenum}

\subsubsection{Overall Comparison}
Figure~\ref{fig:jct-cdf} presents the distributions of JCTs for all four policies, while Table~\ref{tab:exp11} presents makespan and aggregate statistics.
Given the similarities of makespan values between FIFO, SRTF, and FAIR, we can say that \name introduces little overhead.
Furthermore, packing jobs can indeed improve makespan. 
Note that because of online job arrivals, we do not observe large improvement from PACK in this case.
However, when many jobs arrive together, PACK can indeed have a larger impact (\S\ref{sec:eval-automl}).

These experiments also reestablishes the fact that in the presence of known completion times, SRTF can indeed improve the average JCT -- $3.19\times$ w.r.t. FIFO in this case.

\subsubsection{Impact of Fast Job Switching}
We evaluate {\name}'s ability to perform fast job switching in two contexts.
First, we show that it can allow fast preemption, which, in turn, allows us to implement the shortest-remaining-time-first (SRTF) scheduling policy. 
Second, we show that how {\name} can allow seconds-granularity fair sharing between multiple DL jobs -- as opposed to minutes-granularity \cite{gandiva}.
In both cases, we consider a single GPU lane.

\paragraph{SRTF}
Consider the following scenario: a large training job has been running for a while, then the user wants to quickly do some test runs for hyper-parameter tuning for smaller networks. 
Without \name, they would have to wait until the large job finishing -- this is an instance of HOL blocking.
\name enables preemption via efficient switching to run short jobs and resumes the larger job later.

We pick a segment in the long job trace, containing exact the scenario, and record its detailed execution trace, showing in Figure~\ref{fig:srtf-compute}.
When job \#1 arrives, the background job \#0 is immediately stopped and \name switches to run the newly arrived shorter job. Job \#2 comes early than job \#3, but since \#3 is shorter, it is scheduled first. And finally since job \#5
is shorter, \#4 is preempted and let \#5 run to completion. During the process,
the background job \#0 is only scheduled when there is no other shorter job existing.

Figure~\ref{fig:srtf-mem} is another example demonstrating \name's ability to fast switch. It is the visualization of memory allocation activities in the scale of seconds: at the moment of a job switching, the second job's iteration starts immediately after the first job stops.

\begin{figure*}[!t]
    \centering
    \includegraphics{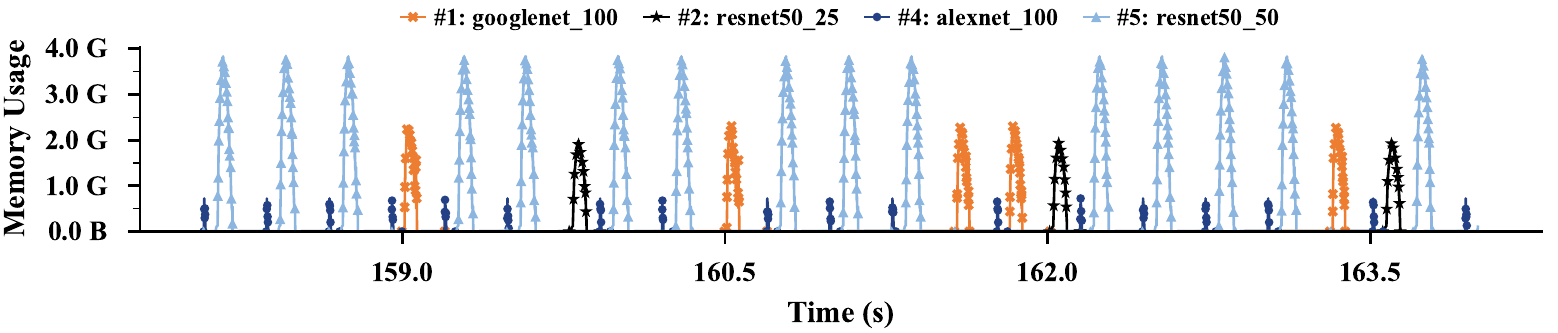}
    \caption{A slice of memory usage for the long trace using FAIR. 
      Note that memory usage does not include actual computation; hence, the temporal gaps.}
    \label{fig:fair}
\end{figure*}

\paragraph{Time Sharing/Fairness}
Figure~\ref{fig:fair} is an snapshot of the job trace running under the FAIR policy. 4 training jobs: \texttt{googlenet\_100}, \texttt{resnet50\_25},
\texttt{alexnet\_100} and \texttt{resnet50\_50} are active during the snapshot,
and \name tries to equalize their GPU time. Again the switches all happen at sub-second granularity.

Note that the depicted memory usage is not a complete representation of the activity of that iteration.
Computation actually continues to run in the gaps.

\begin{figure}[!t]
    \centering
    \includegraphics[]{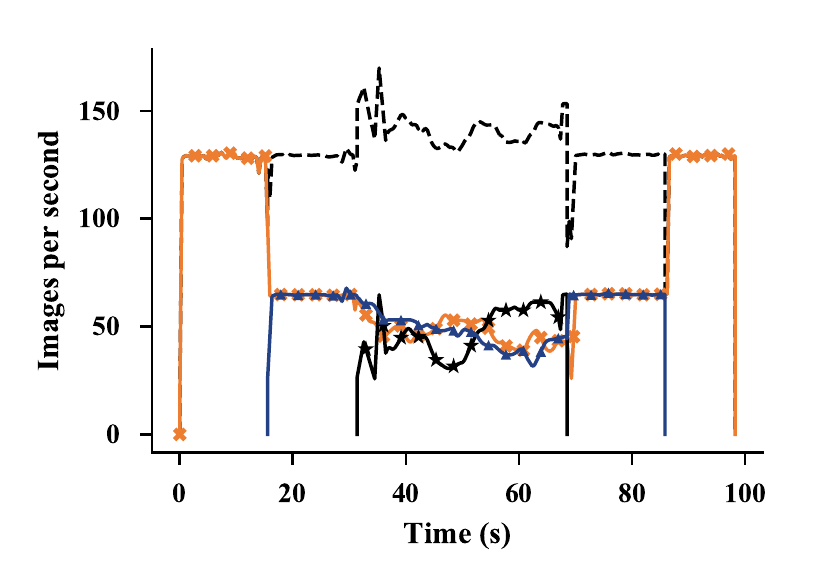}
    \caption{Fair sharing among three jobs, all training \texttt{inception3\_50}. 
      Black dashed line shows the overall throughput.}
    \label{fig:switching-fair}
\end{figure}

To better illustrate the impact of fairness,
we show another microbenchmark, demonstrating \name's ability to switch jobs efficiently using 3 training jobs and focusing on the fair sharing of GPU throughput in Figure~\ref{fig:switching-fair}.

For ease of exposition, we picked three jobs of the same DL model \texttt{inception3\_50} -- this allows us to compare and aggregate training throughput of the three models in terms of images processed per second. 
In this figure, in addition to the throughput of individual jobs, the black dashed line shows the aggregate throughput.

The training jobs start at time 0s, 15s and 30s. 
At 15s, when the second job starts, while the total throughput remains unchanged, each job's share is halved. 
It further reduces to about a third when the third job arrives.
Similarly, the reverse happens when jobs finishes in the reverse order. 
The system throughput roughly remains the same throughout the experiment.
Note that \name reacts almost immediately for job arriving and leaving events.

In contrast, FIFO scheduling or other sharing policies (\eg, MPS) cannot enforce fair sharing. 

\subsection{Hyper-Parameter Exploration}
\label{sec:eval-automl}
Using \name to PACK many jobs is especially useful when many/all jobs are ready to run.
One possible use case for this is automatic hyper-parameter tuning for DL models. 
Typically, hundreds of training jobs are generated in parallel for parameter exploration. 
Most of the generated models will be killed shortly after they are deemed to be of poor quality. 
In this case, increasing the concurrency on GPU can help improve the parameter exploration performance by running multiple small jobs together, whereas today only FIFO is possible.

\begin{figure}[!t]
    \centering
    \includegraphics[scale=0.8]{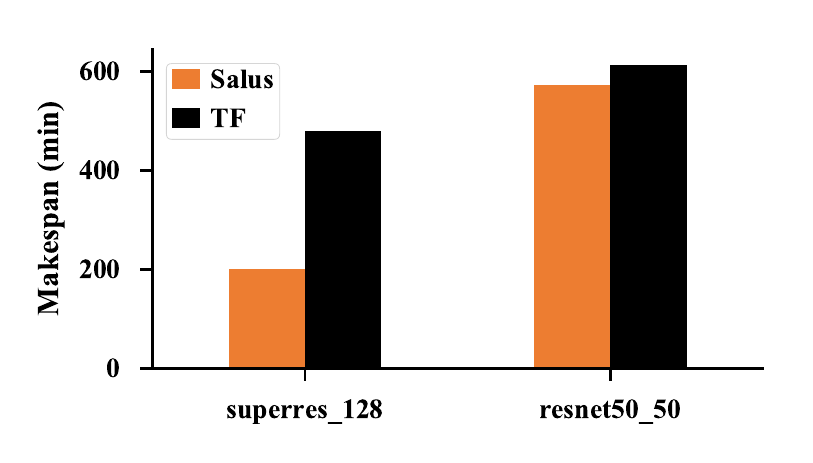}
    \caption{Makespan of two hyper-parameter tuning multi-jobs each of which consists of 300 individual jobs.}
    \label{fig:automl}
\end{figure}

We evaluate two sets of hyper-parameter exploration jobs: \texttt{resnet50\_50} and \texttt{superres\_128}, for image classification and resolution enhancement, respectively. 
Each set has 300 jobs, and each one completes after all 300 complete.
A comparison of achieved makespan using FIFO (in TensorFlow) and \name is shown in Figure~\ref{fig:automl}. In the \texttt{resnet50\_50} case, there is $1.07\times$ makespan improvement while it is $2.38\times$ for \texttt{superres\_128}.

Little improvement is seen for \texttt{resnet50\_50} because even if the GPU has enough memory to hold many of the jobs together, computation likely becomes the bottleneck under such heavy sharing. 
Consequently, the makespan of the whole set of jobs does not see much improvement.

\subsection{Inference}
\label{sec:eval-inference}

\begin{figure*}[!t]
    \centering
    \includegraphics{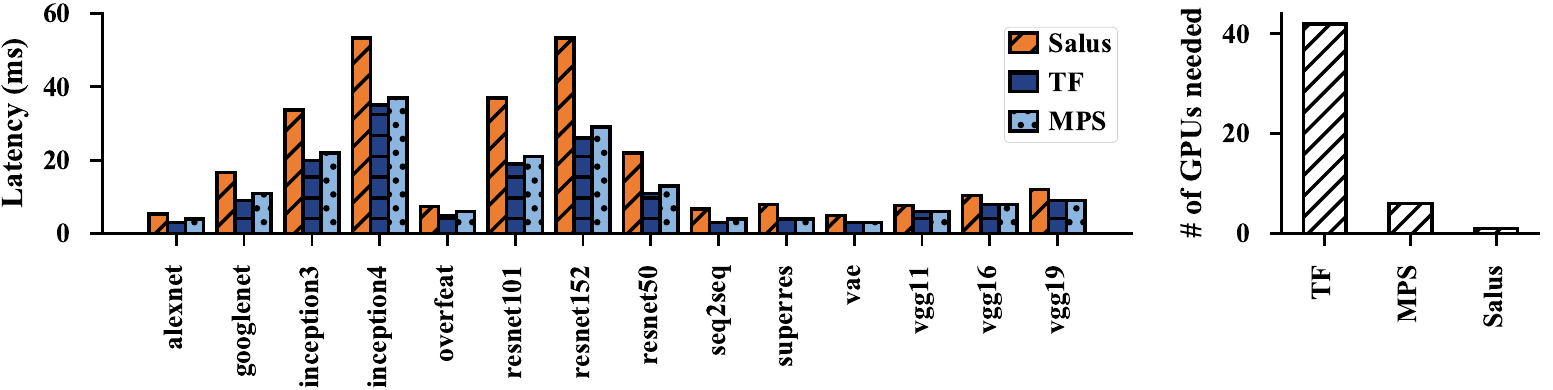}
    \caption{The latencies and number of GPUs needed to host 42 DL models for inference at the same time. 
      3 instances of each network is created. 
      Each model have a low request rate.}
    \label{fig:card270}
\end{figure*}

So far we have only discussed DL training, but we note that serving a trained model, \ie, inference, can also be a good -- if not better -- candidate for GPU memory sharing.
Rather than focusing on throughout when training, latency of individual inference request becomes a more important requirement when serving DL models \cite{clipper,tensorrt}.

In order to keep responsive to requests, DL models have to be online 24x7 hours. 
In the traditional setting, each model must reside on a dedicated GPU. 
However, the traffic of serving requests is not always constant throughout the day, and there are times when
the request rate is significantly lower compared to peak. 
Consolidating DL models into fewer GPUs while remain responsive can save the maintains cost for service providers.

We demonstrate \name's ability to reduce the number of GPUs needed while maintaining reasonable response latency in Figure~\ref{fig:card270}.
42 DL inference jobs are selected consisting of 14 different models, 3 instances for each model.
Without MPS or \name, 42 GPUs are needed to hold these DL models. 
In contrast, \name needs only 1 GPU, achieving $42\times$ utilization improvement, while the average latency overhead is less than 5ms.
For comparison, MPS needs 6 GPUs.

A future work is to detect current request rate for inference jobs and automatically scale up or down horizontally. 
Nevertheless, \name provides the essential primitives that makes the implementation possible.

\subsection{Overhead}
\label{sec:eval-overheads}
\begin{figure}[!t]
    \centering
    \includegraphics[scale=1.2]{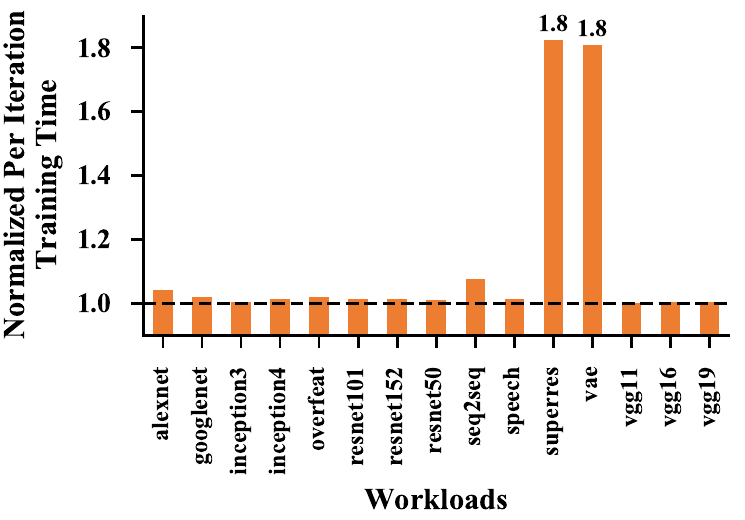}
    \caption{Per iteration time per workload in \name, normalized by that of TensorFlow. Only the largest batch size for each network is shown, as other batch sizes have similar performance.}
    \label{fig:exp5-17}
\end{figure}

\name has to be efficient, otherwise the benefits gained from sharing can be easily offset by the overhead.
Figure~\ref{fig:exp5-17} shows per iteration training time in \name, normalized by per iteration training time in baseline TensorFlow.

For most CNN models, \name has minimal overhead -- less than 10\%, except for a few.
The common point of these high-overhead DL models is that they also performs large portion of CPU computation in addition to heavy GPU usage. 
Since \name implements its own execution engine, the CPU computation is also redirected and sent to \name for execution, which is not yet heavily optimized.

We finally proceed to compare the performance to run two jobs on a single GPU using existing solutions.
We consider the following approaches to enable sharing in addition to using \name:
\begin{denseenum}
    \item \textbf{Static Partitioning (SP)}: Jobs can be executed on the same GPU as long as the sum of two or more consecutive jobs' peak memory do not exceed the device capacity, using non-exclusive mode. 

    \item \textbf{SP + NVIDIA MPS}: NVIDIA MPS is intended to reduce GPU context switch and speed up the execution. 
      It does not manage the GPU memory, so static partitioning is still needed.
    
    \item \textbf{SP + MPS + Overcommit (OC)}: Using the unified memory access and GPU page fault feature introduced in CUDA 8, we can overcommit the GPU memory and let more jobs come in.
\end{denseenum}

\begin{figure}[!t]
    \centering
    \includegraphics{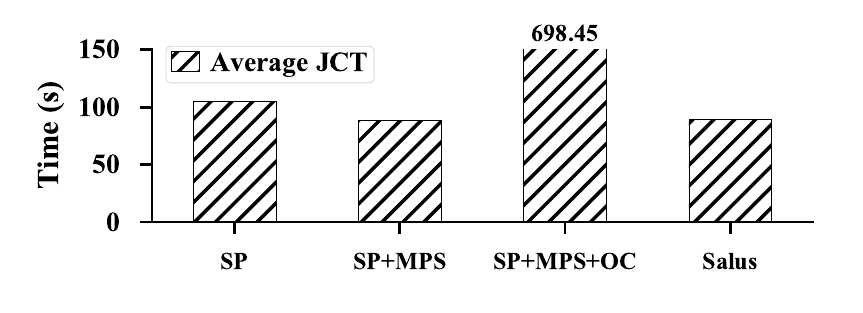}
    \caption{Two concurrent jobs of training \texttt{alexnet\_25} for one minute.}
    \label{fig:exp6_2}
\end{figure}

The setup is simple where two \texttt{alexnet\_25} training jobs are started at the same time and each runs for a minute.
The jobs run on a single GPU made possible using one of the above sharing solution.
We then collect and compare the average JCT and report the result in Figure~\ref{fig:exp6_2}.

The result confirms that MPS is indeed better than SP due to the avoidance of  GPU context switching. Unfortunately, the promising SP+MPS+OC solution has significantly bad performance that is beyond useful at the moment. 
\name manages to achieve almost the same performance as MPS while providing much more flexibility in scheduling policy.
As shown before, in lightly-loaded inference scenarios, it can significantly outperform MPS in terms of utilization.

\section{Related Work}

\paragraph{Container-Based Sharing}
At the largest granularity, container-based solutions have been adopted by modern cluster managers such as Apache YARN~\cite{yarn}, Mesos~\cite{mesos}, and Kubernetes~\cite{kubernetes}.
They typically treat GPUs at device-granularity; \ie, an allocation consists of one or more whole GPUs.
Clearly, the problem of underutilization \textit{per-gpu} remains.

Recent work on nvidia-docker~\cite{nvidiaDocker} makes using one NVIDIA GPU in multiple containers possible, but it is
essientially the same as removing exclusive-mode on GPU.
As already discussed, this is not ideal and has performance issues.
It is possible to use NVIDIA MPS, but the support is not complete yet~\cite{nvidiaDocker-issue419}.

Therefore, {\name} is complementary to these work and can be used to enable fractional GPU allocations.

\paragraph{GPU Virtualization via Library Interception}
To share one single GPU among applications, virtualization via library interception or API remoting is a popular trick to bypass default hardware/driver implementations.
Examples are gVirtuS \cite{gvirtus}, GViM \cite{gvim}, vCUDA \cite{vcuda}, rCUDA \cite{rcuda}, and work from Ravi {\etal} \cite{cloud-gpu-sharing} to share GPU in cloud environments.
However, most of the work focus on sharing in terms of GPU applications of no more than a few
kernels. 
Modern DL applications usually make use of hundreds of unique GPU kernels during their training, and they also rely on advanced CUDA APIs (\eg, CUDA stream callbacks) that are often not supported in these works.

In addition, as an official implementation with similar techniques, CUDA MPS lacks wide support for DL frameworks. 
For example, TensorFlow crashes when running two instances on the MPS Server \cite{tf-issue-9080}.
Gandiva \cite{gandiva} uses MPS as well, and the authors report unpredictable performance. 

\paragraph{New API}
Instead of hacking library API, others choose to create new sets of API from scratch; \eg, Pagoda~\cite{pagoda}, GeMTC~\cite{gemtc}, \etc. 
While this achieves the most flexibility and efficiency, it is in practice hard if not impossible for existing DL frameworks to be adapted to the new API.

\paragraph{Efforts for Increasing GPU Utilization}
Rather than packing more applications into one GPU, another completely different approach to increase GPU utilization focuses on single application use cases.
Some attempts to statically fuse multiple tasks together to increase efficiency; \eg, TensorFlow XLA~\cite{tensorflow-xla}, NNVM~\cite{mxnet}, and \cite{fine-grained-gpgpu}.
While other works focus on GPU kernel level concurrency and scheduling~\cite{gpgpu-concurrency}.
{\name} is complementary to these approaches. 

\section{Concluding Remarks}
GPUs have emerged as the primary computation devices for deep learning (DL) applications. 
However, modern GPUs and their runtimes do not allow multiple processes to coexist in a GPU. 
As a result, unused memory of a DL job remains unaccessible to other jobs, leading to large efficiency, performance loss, and head-of-line (HOL) blocking.

{\name} is a consolidated execution service that enables fine-grained GPU sharing between complex, unmodified DL jobs. 
It achieves this by exposing two important primitives: 
(1) \emph{fast job switching} that can be used to implement time sharing and preemption; and 
(2) the \emph{GPU lane} abstraction to enable dynamic memory partitioning, which can be used for packing multiple jobs on the same GPU.
Together, they can be used to implement unforeseen scheduling policies as well. 
Our integration of {\name} with TensorFlow shows that {\name} can allow multiple DL jobs to coexist, enable fair sharing and preemption between them, and improve overall efficiency and DL training performance in a shared-GPU environment.

However, {\name} is only a first attempt, and it opens many interesting research challenges.
First and forement, {\name} provides a mechanism but the question of policy -- what is the best scheduling algorithm for DL jobs running on a shared GPU? -- remains open. 
Second, while not highlighted in the paper, {\name} can be extended to multiple GPUs or even other accelerators on the same machine.
Finally, we plan to extend it to GPUs across multiple machines leveraging RDMA.

\label{EndOfPaper}

\bibliographystyle{abbrv}
\bibliography{salus-arxiv}

\clearpage

\section*{Appendix}

\subsection{Workloads}
Table~\ref{tab:workloads} is the full list of workloads and their batch sizes we used
in our evaluation.

\begin{figure}[b]
    \centering
    \includegraphics[width=\columnwidth]{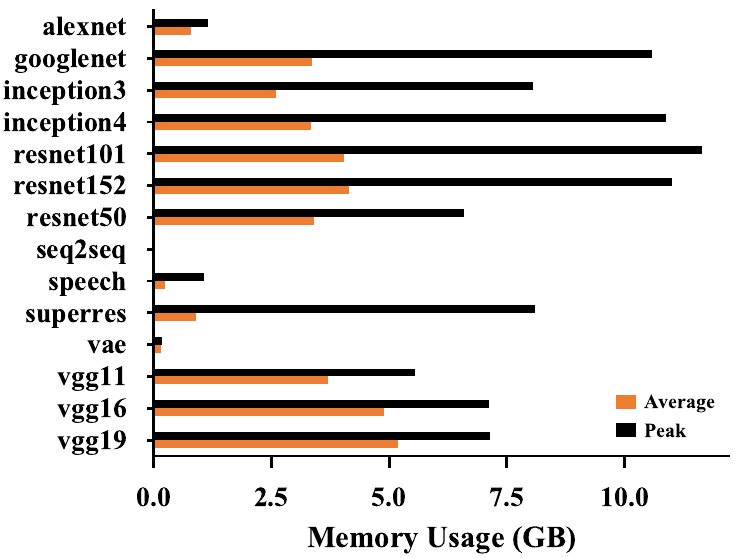}
    \caption{Average and peak GPU memory usage per workload, measured in PyTorch 
    and running on NVIDIA P100 with 16 GB memory.
    The average and peak usage for vae is 156 MB, 185 MB, which are too small to show in the figure.
    }
    \label{fig:mem-pytorch}
\end{figure}

Figure~\ref{fig:mem-pytorch} is the same peak and average GPU memory usage measurement
done in PyTorch, except \texttt{overfeat}, which we could not find a working implementation.

\begin{table}[h!]
\small
\centering
\begin{tabular}{lll}
    \toprule
    Model                 & Type          & Batch Sizes\\
    \midrule
     \texttt{alexnet}     & Classification               & 25, 50, 100\\
     \texttt{googlenet}   & Classification               & 25, 50, 100\\
     \texttt{inception3}  & Classification               & 25, 50, 100\\
     \texttt{inception4}  & Classification               & 25, 50, 75\\
     \texttt{overfeat}    & Classification               & 25, 50, 100\\
     \texttt{resnet50}    & Classification               & 25, 50, 75\\
     \texttt{resnet101}   & Classification               & 25, 50, 75\\
     \texttt{resnet152}   & Classification               & 25, 50, 75\\
     \texttt{vgg11}       & Classification               & 25, 50, 100\\
     \texttt{vgg16}       & Classification               & 25, 50, 100\\
     \texttt{vgg19}       & Classification               & 25, 50, 100\\
     \texttt{vae}         & Auto Encoder                 & 64, 128, 256 \\
     \texttt{superres}    & Super Resolution             & 32, 64, 128 \\
     \texttt{speech}      & NLP                          & 25, 50, 75 \\
     \texttt{seq2seq}     & NLP                          & Small, Medium, Large \\
    \bottomrule
\end{tabular}
    \caption{DL models, their types, and the batch sizes we used.
      Note that the entire network must reside in GPU memory when it is running.
      This restricts the maximum batch size we can use for each network.}
\label{tab:workloads}
\end{table}

\end{document}